\documentclass[12pt,times]{elsarticle}
\usepackage[T1]{fontenc}
\usepackage{amsbsy}
\usepackage[colorlinks]{hyperref}
\usepackage{amsmath}
\usepackage{color}
\usepackage{multirow}
\usepackage{booktabs}
\usepackage{subfig}
\usepackage{float}
\biboptions{authoryear}
\begin{document}
\begin{frontmatter}
\title{An additive Mori-Tanaka scheme for elastic-viscoplastic composites based on a modified tangent linearization}
\author[1]{K. Kowalczyk-Gajewska\corref{mycorrespondingauthor}}
 \ead{kkowalcz@ippt.pan.pl}
\cortext[mycorrespondingauthor]{Corresponding author, fax: +4822 8269815}
\author[2]{S. Berbenni\corref{mycorrespondingauthor}}
\ead{stephane.berbenni@univ-lorraine.fr}
\author[2]{S. Mercier}
\address[1]{Institute of Fundamental Technological Research, Polish Academy of Sciences,\\
Pawi\'{n}skiego 5B, 02-106 Warsaw, Poland}
\address[2]{Universit{\'{e}} de Lorraine, Arts et Métiers Paris Tech, CNRS, LEM3, F-57000 Metz, France}

\begin{abstract}
Mean-field modelling based on the Eshelby inclusion problem poses some difficulties when the non-linear Maxwell-type constitutive law is used for elasto-viscoplasticity. One difficulty is that this behavior involves different orders of time differentiation, which leads a long-term memory effect. One of the possible solutions to this problem is the additive interaction law. 
Generally, mean field models solely use the mean values of stress and strain fields per phase, while variational approaches consider the second moments of stresses and strains. It is seen that the latter approach improves model predictions allowing to account for stress fluctuation within the phases. However, the complexity of the variational formulations still makes them difficult to apply in the large scale finite element calculations and for non-proportional loadings. Thus, there is a need to include the second moments within homogenization models based on the additive interaction law. In the present study, the incorporation of the second moments of stresses into the formulation of the additive Mori-Tanaka model of two-phase elastic-viscoplastic material is discussed. A modified tangent linearization of the viscoplastic law is proposed, while the Hill-Mandel's lemma is used to track the evolution of second moments of stresses.  To study the model performance and efficiency, the results are compared to the full-field numerical calculations and predictions of other models available in the literature. Very good performance of the modified tangent linearization is demonstrated from these benchmarks for both monotonic and non monotonic loading responses.
\end{abstract}
\end{frontmatter}

\section{Introduction}
\label{sec1}

{
Mean-field {micromechanical modeling} based on the Eshelby result poses some difficulties when the non-linear Maxwell-type constitutive law is used for elasto-viscoplasticity. First, it involves different orders of time differentiation, which leads to long-term memory effect so that an aggregate of Maxwellian constituents is no more Maxwellian \citep{Suq87}. Second, the adaptation of linear homogenization methods to such non-linear behavior still needs some refinements to reach enough predictive capability as compared to full-field calculations. One of the possible solutions to this problem is the additive interaction law \citep{Molinari97,Molinari02,Mercier05,Mercier09,Mercier12}. {The law is named 'additive' as \cite{Molinari97} proposed to add the elastic and viscoplastic interaction laws to approximate the Eshelby solution for elastic-viscoplastic materials, following the additive split of the strain rate tensor valid for the Maxwellian constituents. It is interesting to note that the Mori-Tanaka averaging scheme based on the sequential linearization
method of \citep{Kowalczyk11a} is equivalent to the one developed by the previous authors based on the additive interaction laws. The additive law should not be confused with the differential (or incremental) schemes in the frame of which the effective properties of the composite are obtained by a gradual addition of infinitesimal (or incremental) quantities of reinforcements, (see \cite{McLaughlin77,Broohm00,Dhar23}).} Numerical results of the additive interaction law for non-linear particulate elasto-viscoplastic composites were compared to full-field finite element (FEM) calculations in \cite{Czarnota15} and \cite{Mercier19}. Alternative proposals are based e.g. on the translated fields method \citep{Paquin99,Sabar02,BerbenniCapolungo15,Mareau15,Lhadi18,Tsekpuia23}, {on the use of the Laplace-Carson transform and solving the thermoelastic-like problem in the Laplace-Carson transformed space, see e.g.: \cite{Hashin69,Laws78,Masson99,Brenner05,Pierard06,Ricaud09,Masson20} or on the variational approaches \citep{Lahellec07,Lahellec13,Brassart2012,Badulescu15,Agoras16,Cotelo20,Das21}.
}

{
{The additive law and the translated field method} use solely the mean values of stress and strain fields per phase, while the second moments of stresses were considered in variational approaches. It is seen that the latter approach improves model predictions allowing to account for stress fluctuation within the phases {\citep{Lahellec13,Badulescu15,Cotelo20}}. Then, there is a need to include the second moments of stresses within the remaining direct approaches. A pioneering work in this area was Suquet’s modified secant method \citep{Suquet95}, which was restricted to monotonic loadings. {For non-linear purely viscoplastic composites, a modified affine formulation was proposed by \cite{Brenner01}, using the second moments of stresses in the tangent viscoplastic compliance tensor of the phases}. For non-monotonic loadings, \cite{Doghri2011} and \cite{Wu15} respectively developed an incremental-tangent method and an incremental-secant procedure for elasto-plastic composites, where the second moments are calculated based on an elastic trial state. Later, \cite{Wu17} extended the incremental-secant procedure to elasto-viscoplastic composites. More
recently, \cite{Masson20} and \cite{Berbenni21} developed independently a  modified secant linearization of the viscoplastic strain rate to extend direct approaches for two-phase elasto-viscoplastic composites. In \cite{Masson20}, a Mori-Tanaka homogenization scheme was derived using the correspondence principle \citep{Man66,Hashin69,Laws78} based on the collocation method to invert the Laplace–Carson transforms as first reported in \cite{Ricaud09}. {Let us note that, in the similar context, the second moments of stresses were earlier calculated in the Laplace–Carson space for application to non-linear viscoelastic composites by \cite{Brenner05} in the framework of the hereditary affine approach of \cite{Masson99}. However, it was only developed for specific loading paths like creep tests, using a “quasi-elastic” approximation for Laplace-Carson inversion to directly derive the material response in the real time space.} In \cite{Berbenni21}, a Mori-Tanaka homogenization scheme was derived based on the exact solution of viscoelastic Eshelby ellipsoidal inclusion
problem in the time domain as first reported in \cite{Berbenni15}. The additional relation for tracking the second moments of stresses in the matrix phase was derived from the Hill-Mandel's lemma \citep{Hill67}. It was earlier used by \cite{LiWeng97} for compressible two-phase elasto-vicoplastic composites under creep loading, and by \cite{Badulescu15} for incompressible two-phase linear viscoelastic composites. Very recently, \cite{Lahellec24} used both the effective free energy and the effective energy dissipated in linearly viscoelastic particulate composites to derive two relations that extend the result of \cite{Badulescu15} to derive the second moments of stresses.
}

In the present study, the incorporation of the second moments of stresses into the formulation of the additive Mori-Tanaka (MT) model of two-phase elastic-viscoplastic materials is developed. In contrast with \cite{Masson20} and \cite{Berbenni21} who considered a modified secant formulation, a \emph{modified tangent} linearization of the viscoplastic law is performed for the first time in the framework of the additive MT model, while similarly the Hill-Mandel's lemma is used to track the evolution of second moments of stresses. To study the model performance the results are compared to the full-field numerical calculations and predictions of other models available in the literature. Very good performance of the new linearization scheme is demonstrated from these benchmarks.

{
The paper consists of five sections. After the present introductory part, the formulation of the additive Mori-Tanaka model for elastic-viscoplastic composite material is presented in Section \ref{sec2}.  Section \ref{Sec:Lin} discusses two variants of the tangent linearization of the viscoplastic matrix response required to apply the Eshelby's results: the standard one employing solely mean values (first moments) of stress and strain in the phases and the modified one, which is based on the second moments of stresses. The present developments are applied in the case of  isotropic elasticity and Perzyna-type viscoplasticity as illustrations. Let us note that as our theory is general, other constitutive model can be adopted. In Section \ref{Sec:Results}, a validation of the new modified tangent linearization scheme is performed using the reference full-field results by fast Fourier transform (FFT) method found in the literature. The predictions are also compared to those of the standard additive MT model with classic tangent linearization. {After the conclusion section, two appendices are proposed, which} contain details of solution algorithms and comparison of the proposed model with the modified secant approaches, respectively.
}

\section{Mori-Tanaka scheme based on the additive interaction law}
\label{sec2}
{\subsection{Notations and conventions}
\label{sec21}}
By default the so-called absolute notation is used. When an index notation is used the Einstein summation convention on repeated indices applies. Scalar, second order tensor and fourth order tensor variables are denoted using italic $a$, boldface $\mathbf{A}, \boldsymbol{\alpha}$ and blackboard $\mathbb{A}$ characters, respectively. A full contraction of two tensors of two arbitrary orders is denoted by $\cdot$, namely: $\mathbb{A}\cdot\mathbf{B}$, $\mathbf{A}\cdot\mathbf{B}$ (in an index notation:  $A_{ijkl}A_{kl}$, $A_{kl}B_{kl}$), while contraction of two tensors of the same order resulting in the tensor of the same rank is denoted with no sign, i.e $\mathbf{A}\mathbf{B}=\mathbf{C}$ or $\mathbb{A}\mathbb{B}=\mathbb{C}$ (in an index notation: $A_{ik}B_{kj}=C_{ij}$ or $A_{ijkl}B_{klmn}=C_{ijmn}$). The external product of two tensors is denoted by $\otimes$, i.e. $\mathbf{A}\otimes\mathbf{B}=\mathbb{C}$ (in index notation $A_{ij}B_{kl}=C_{ijkl}$). The second order and symmetric fourth order identity tensors are denoted by $\mathbf{I}$ (with components given by the Kronecker delta $\delta_{ij}$) and $\mathbb{I}$
(with components $1/2(\delta_{ik}\delta_{jl}+\delta_{il}\delta_{jk})$), respectively. A deviator $\mathbf{a}$ of the second order tensor $\boldsymbol{\alpha}$ is calculated as
$\mathbf{a}=\boldsymbol{\alpha}-1/3(\mathrm{tr}\boldsymbol{\alpha})\mathbf{I}$, where trace of the tensor means $\mathrm{tr}\boldsymbol{\alpha}=\alpha_{kk}$. Fourth order tensors $\mathbb{I}_{\rm{P}}$ and $\mathbb{I}_{\rm{D}}$ are the orthogonal projectors such that $\mathbb{I}_{\rm{P}}\cdot\boldsymbol{\alpha}=1/3(\mathrm{tr}\boldsymbol{\alpha})\mathbf{I}$ and $\mathbb{I}_{\rm{D}}\cdot\boldsymbol{\alpha}=\mathbf{a}$ (deviator of $\boldsymbol{\alpha}$) where $\mathbb{I}_{\rm{P}}=1/3\mathbf{I}\otimes\mathbf{I}$ and $\mathbb{I}_{\rm{D}}=\mathbb{I}-\mathbb{I}_{\rm{P}}$. {The volume average of some field over the phase $k$ ($k=i$ for the inclusion phase and $k=m$ for the matrix phase) is denoted by $\overline{(.)}_k$ and is defined as the integral: $\overline{(.)}_k=1/V_k\int_{V_k}( . )dV$.}

\subsection{Position of the problem} 
In the present work, we investigate the homogenized behavior of a two-phase particulate composite, {however,} the theory is valid for particulate material with several inclusion phases.  Each phase has an elastic-viscoplastic behavior whose response is assumed of Maxwell type. %
So, the total strain rate tensor is additively decomposed into elastic and viscoplastic parts {(assuming a small strain setting)}:
\begin{equation}
	\dot{\boldsymbol{\varepsilon}}=\dot{\boldsymbol{\varepsilon}}^e+\dot{\boldsymbol{\varepsilon}}^v
\end{equation}
Linear elasticity is considered while {a non linear viscoplastic model through tensorial function $\bf{g}(\boldsymbol{\sigma})$} is accounted for:
\begin{equation}\label{Eq:nonlinev}
	\dot{\boldsymbol{\varepsilon}}^e=\mathbb{M}^e\cdot\dot{\boldsymbol{\sigma}}\,,\quad 	\dot{\boldsymbol{\varepsilon}}^v=\mathbf{g}(\boldsymbol{\sigma})
\end{equation}
Similarly to \cite{Czarnota15} and \cite{Berbenni21}, 
a Perzyna-type {viscoplastic} model {is adopted for $\bf{g}(\boldsymbol{\sigma})$} with
an over-stress function $\Phi$ following \cite{Perzyna86} :
\begin{equation}\label{Eq:Perzyna}
\dot{\boldsymbol{\varepsilon}}^v=\mathbf{g}(\boldsymbol{\sigma})=\Phi \frac{\partial f}{\partial
\boldsymbol{\sigma}}\,
\end{equation}
with
\begin{equation}\label{EQ:Perzynaf}
f=\sigma_{eq}-\sigma_Y-R(\varepsilon_{eq})\,,\quad
\Phi=\dot{\varepsilon}_0\left(\frac{\left<\,f\,\right>}{\sigma_Y+R(\varepsilon_{eq})}\right)^{\frac{1}{M}}
\end{equation}
{and the Macaulay brackets $\left<\, . \,\right>$ denoting a ramp function, i.e. $\left<\,x\,\right>=x$ if $x>0$ and $\left<\,x\,\right>=0$ otherwise.}
In Eqs. \eqref{Eq:Perzyna} and \eqref{EQ:Perzynaf},  $\sigma_{eq}= \sqrt{3/2
\mathbf{s} \cdot \mathbf{s}}$  is the
equivalent Huber-von Mises stress,   $\mathbf{s}$ being  {the deviator of stress tensor}.  Then, $\dfrac{\partial f}{\partial
\boldsymbol{\sigma}}=\dfrac{3 \mathbf{s}}{2\sigma_{eq}}$. The
function $R(\varepsilon^{eq})$  describes the
hardening behavior of the phase. $\varepsilon_{eq} = \int  \dot{\varepsilon}_{eq} dt $ is the {cumulated} plastic strain where 
$\dot{\varepsilon}_{eq}\equiv\sqrt{2/3
\dot{\boldsymbol{\varepsilon}}^v \cdot
\dot{\boldsymbol{\varepsilon}}^v}$
is the equivalent viscoplastic strain rate.

In the overstress model described in Eq. \eqref{EQ:Perzynaf}, the viscoplastic flow is initiated when the function
$f$ is positive. Both parameters $M$ (strain rate sensitivity) and $\dot{\varepsilon}_0$ ({reference strain rate}) represent the rate dependence of the viscoplastic flow {rule}, and $\sigma_Y$ is the  initial flow stress under static conditions. The non-linearity of the viscoplastic flow rule is defined with the power law exponent $1/M$. While our approach can manage {strain-hardened materials}, all the presented results are obtained without hardening, i.e. $R(\varepsilon_{eq})=0$.
With this assumption, the Perzyna-type model can
be rewritten in the following scalar form, which is only valid after initiation of the viscoplastic flow:
\begin{equation}\label{Eq:Perzyna-seq}
\sigma_{eq}=\sigma_Y\left(1+\left(\frac{\dot{\varepsilon}_{eq}}{\dot{\varepsilon}_0}\right)^M\right)\,.
\end{equation}
 {Starting from Eqs. \eqref{Eq:Perzyna} and \eqref{EQ:Perzynaf} and assuming $R(\varepsilon_{eq})=0$, an alternative convenient form for the non-linear viscoplastic strain rate tensor is given by}:
\begin{equation}\label{Eq:NortonPerzyna-N}
	\dot{\boldsymbol{\varepsilon}}^v=\sqrt{\frac{3}{2}}\dot{\varepsilon}_{0}\left(\frac{\left<\sigma_{eq}-\sigma_Y\right>}{\sigma_Y}\right)^{\frac{1}{M}}\mathbf{N}_s
\end{equation}
with
\begin{equation}
	 \mathbf{N}_s=\mathbf{s}/\sqrt{\mathbf{s}\cdot\mathbf{s}}\quad(\mathbf{N}_s\cdot\mathbf{N}_s=1)
\end{equation}

Since we focus on the introduction of second moment of stresses in the Mori-Tanaka {(MT)} scheme based on the tangent additive interaction law, it seems important to recall briefly the {framework} of the original MT scheme, {based on the additive interaction law with first order moments of stresses} \citep{Mercier09,Kowalczyk11a,Mercier12}.

\subsection{Additive interaction law}\label{sec23}

The tangent additive interaction law has been proposed by \cite{Molinari02} as an approximate solution of the Eshelby problem when an elastic-viscoplastic ellipsoidal inclusion is embedded in an infinite  elastic-viscoplastic reference medium. In our work,  the shape of inclusions is  spherical. The theory is more general and is also valid for ellipsoidal inclusions.

  As proposed by \cite{Mercier09}, when the overall response of the particulate composite is modeled by a Mori-Tanaka scheme, the mean strain rate in the inclusion phase  $\dot{\bar{\boldsymbol{\varepsilon}}}_i$  can be linked to the mean  strain rate in the matrix $\dot{\bar{\boldsymbol{\varepsilon}}}_m$ according to tangent additive interaction law:
\begin{equation}\label{Eq:add-law}
	\dot{\bar{\boldsymbol{\varepsilon}}}_i-\dot{\bar{\boldsymbol{\varepsilon}}}_m=-\mathbb{M}_e^*\cdot(\dot{\bar{\boldsymbol{\sigma}}}_i-	\dot{\bar{\boldsymbol{\sigma}}}_m)-\mathbb{M}_v^*\cdot(\bar{\boldsymbol{\sigma}}_i-	\bar{\boldsymbol{\sigma}}_m)
\end{equation}
where $\dot{\bar{\boldsymbol{\sigma}}}_k$ and ${\bar{\boldsymbol{\sigma}}}_k$ ($k=i,m$) are mean stress rate and stress per phase (first moments). Fourth order tensors  $\mathbb{M}_e^*$ and $\mathbb{M}_v^*$ are inverse Hill tensors corresponding to the purely elastic and purely viscous solution of the Eshelby-type problem. 

{{The additive interaction law \eqref{Eq:add-law} can be rationalized based on several arguments. \cite{Hashin69} found
an analytical solution for the linear viscoelastic problem for a spherical inclusion embedded in an infinite matrix using Laplace transforms (i.e. correspondence principle). The material response was of the Maxwell type in both phases and incompressibility was assumed. This solution can be exactly retrieved by using the additive interaction law, see \citep{Kouddane93} for a simplified version of Eq. \eqref{Eq:add-law}. The additive interaction law was later applied for solving elasto-viscoplastic matrix-inclusion problems, \citep{Molinari97,Molinari02}. According to the literature \citep{Hill65,Kouddane93} the instantaneous elastic response due to a jump in the external loading appears to be well restituted by the classical interaction law below
\begin{equation}
\dot{\bar{\boldsymbol{\varepsilon}}}_i^e-\dot{\bar{\boldsymbol{\varepsilon}}}_m^e=-\mathbb{M}_e^*\cdot(\dot{\bar{\boldsymbol{\sigma}}}_i-	\dot{\bar{\boldsymbol{\sigma}}}_m)
\end{equation}
For a quasi-nonlinear creep viscoplastic response, the tangent interaction law proposed by \cite{Molinari87} and adopted in the VPSC software \citep{Lebensohn93} is also providing consistent predictions:
\begin{equation}
\dot{\bar{\boldsymbol{\varepsilon}}}_i^v-\dot{\bar{\boldsymbol{\varepsilon}}}_m^v=-\mathbb{M}_v^*\cdot(\bar{\boldsymbol{\sigma}}_i-	\bar{\boldsymbol{\sigma}}_m)
 \end{equation}
From this observation, \cite{Molinari97} have proposed to add the two previous interaction relations. With that, one recovers Eq. \eqref{Eq:add-law}.  
It was verified by \cite{Mercier05} that, for elastic-viscoplastic materials, the predictions of the additive interaction law for the single inclusion problem were in good agreement with Finite Element simulations for various loading conditions and strain paths.}}

Let us note that the inverse Hill tensor \citep{Hill65} is calculated as:
\begin{equation}\label{Eq:MHilldef}
	\mathbb{M}^*=(\mathbb{L}^*)^{-1}=(\mathbb{P}^{-1}-\mathbb{L})^{-1}\,\quad \mathbb{S}=\mathbb{P}\mathbb{L},\quad  \mathbb{L}=\mathbb{M}^{-1}
\end{equation}
where $\mathbb{P}$ is called the polarisation tensor in \citep{Hill65} and $\mathbb{S}$ is the Eshelby tensor. 
The tensors $\mathbb{P}(\mathbb{L})$ depend on the shape of inclusions and the matrix mechanical response (elastic or viscoplastic). 

{In our work, the elastic behavior of the matrix phase, i.e. the reference medium for the MT scheme, is supposed to be linear and isotropic, hence $\mathbb{L}_e=3K\mathbb{I}_{\rm{P}}+2\mu\mathbb{I}_{\rm{D}}$,
where $K$ and $\mu$ are bulk and shear elastic moduli of the matrix, and, $\mathbb{I}_{\rm{P}}$ and $\mathbb{I}_{\rm{D}}$ were defined in section \ref{sec21}. For spherical inclusion, the exact solution of $\mathbb{P}_e(\mathbb{L}_e)$ for elasticity is given in a closed form by:}
\begin{equation}
\mathbb{P}_e(\mathbb{L}_e)=\frac{1}{3K+4\mu}\mathbb{I}_{\rm{P}}+\frac{3K+6\mu}{5\mu(3K+4\mu)}\mathbb{I}_{\rm{D}}
\end{equation} 
 Then the inverse Hill tensor for elasticity specifies as:
\begin{equation}\label{Eq:MHill}
\mathbb{M}^*_e=\frac{1}{4\mu}\mathbb{I}_{\rm{P}}+\frac{1}{2\mu}\frac{2(3K+6\mu)}{9K+8\mu}\mathbb{I}_{\rm{D}}\,.
\end{equation}
The polarization tensor $\mathbb{P}_{v}$ associated to the viscoplastic flow is more complex to determine. Indeed,  the mathematical form of the viscous stiffness tensor depends on the applied linearization method, {e.g. secant vs.  tangent moduli or compliances.}
 This brings us to the linearization issue for the non-linear {viscoplastic constitutive equation} \eqref{Eq:nonlinev} of the matrix material. Next, we will {first} recall the original tangent linearization procedure proposed by \cite {Molinari02} or \cite{Molinari87}, which employs only first moments, see Section \ref{sec31}. {Second, the} novelty of the present work will be detailed in Section \ref{sec32}. Applying the Hill-Mandel's lemma following \cite{Masson20} and \cite{Berbenni21}, a new modified tangent linearization procedure with second moments of stresses is derived.


\section{Modified tangent linearization with
second moments of stresses}\label{Sec:Lin}

\subsection{Original tangent linearization using first moments of stresses} 
\label{sec31}

{The local viscoplastic strain rate response given in Eq. \eqref{Eq:nonlinev} is first linearized using the tangent viscoplastic compliance tensor $\mathbb{M}^{v-tg}(\boldsymbol{{\sigma}})=\frac{\partial \mathbf{g}(\boldsymbol{\sigma})}{\partial{\boldsymbol{\sigma}}}$ and a back-extrapolated strain rate tensor $\dot{\boldsymbol{\varepsilon}}^{v-ref}(\boldsymbol{{\sigma}})$ as follows:
\begin{equation}\label{Eq:Affine}
	\dot{\boldsymbol{\varepsilon}}^v
=\mathbb{M}^{v-tg}(\boldsymbol{{\sigma}})\cdot\boldsymbol{\sigma}+\dot{\boldsymbol{\varepsilon}}^{v-ref}(\boldsymbol{{\sigma}})\,.
\end{equation} 
To develop the tangent additive interaction law, it has been proposed that Eq. \eqref{Eq:Affine} 
is approximated with a fictitious linear thermo-elastic material using first moments of stress tensor, i.e. 
\begin{equation}\label{Eq:Affine-First}
	\dot{\boldsymbol{\varepsilon}}^v
=\mathbb{M}^{v-tg}(\bar{\boldsymbol{{\sigma}}})\cdot\boldsymbol{\sigma}+\dot{\boldsymbol{\varepsilon}}^{v-ref}(\bar{\boldsymbol{{\sigma}}})\,.
\end{equation} 
In the case of the Perzyna-type law \eqref{Eq:NortonPerzyna-N}
\begin{equation}
\label{Eq:Mvtg} 
\mathbb{M}^{v-tg}(\boldsymbol{{\sigma}})\approx\mathbb{M}^{v-tg}(\bar{\boldsymbol{{\sigma}}})=\mathbb{M}^{v-tg}(\bar{\mathbf{N}}_s,\bar{\sigma}_{eq})=\frac{1}{2\mu^{tg}({\bar{\sigma}_{eq}})}
\bar{\mathbf{N}}_s \otimes \bar{\mathbf{N}}_s +\frac{1}{2\mu^{sec}({\bar{\sigma}_{eq}})}(\mathbb{I}_{\rm{D}}-\bar{\mathbf{N}}_s\otimes \bar{\mathbf{N}}_s)
\end{equation}
where
\begin{equation}\label{Eq:moduli}
2\mu^{tg}(\bar{\sigma}_{eq})=\frac{2M\sigma_Y}{3\dot{\varepsilon}_0}
\left(\frac{\left<\,\bar{\sigma}_{eq}-\sigma_Y\,\right>}{\sigma_Y}\right)^{1-\frac{1}{M}} ,\quad 2\mu^{sec}(\bar{\sigma}_{eq})=\frac{2\bar{\sigma}_{eq}}{3\dot{\varepsilon}_0}
\left(\frac{\sigma_Y}{\left<\,\bar{\sigma}_{eq}-\sigma_Y\,\right>}\right)^{\frac{1}{M}}.
\end{equation}}
With the above definitions, combining Eqs \eqref{Eq:NortonPerzyna-N}, \eqref{Eq:Affine} and \eqref{Eq:Mvtg}, it is straightforward to obtain the back extrapolated strain rate {depending on the first moments of stresses}:
\begin{equation}\label{Eq:epref}
    \dot{\boldsymbol{{\varepsilon}}}^{v-ref}(\boldsymbol{{\bar{\sigma}}})=\dot{\boldsymbol{\varepsilon}}^{v-ref}(\bar{\sigma}_{eq},\bar{\mathbf{s}})=\left(\frac{1}{2\mu^{sec}(\bar{\sigma}_{eq})}-\frac{1}{2\mu^{tg}(\bar{\sigma}_{eq})}\right)\bar{\mathbf{{s}}}\,.
\end{equation}

Following the MT procedure of mean-field homogenization, it is proposed that $\mathbb{P}_v$ is calculated using the viscous tangent compliance for the matrix phase evaluated using the mean stress tensor in the matrix. Thus, using the subscript $m$ for the matrix phase in the above formula, this tensor reads:
\begin{equation}\label{Eq:tan-first}
\mathbb{M}^{v-tg}(\bar{\boldsymbol{\sigma}}_m)=\mathbb{M}^{v-tg}(\bar{\mathbf{N}}_s,\bar{{\sigma}}_{eq})\,\quad\textrm{where}\quad
\bar{\sigma}_{eq}=\sqrt{3/2\bar{\mathbf{s}}_m\cdot\bar{\mathbf{s}}_m}\,,\quad \bar{\mathbf{N}}_s=\bar{\mathbf{s}}_m/\sqrt{\bar{\mathbf{s}}_m\cdot\bar{\mathbf{s}}_m}
\end{equation}
where $\bar{\mathbf{s}}_m$ is the deviatoric part of $\bar{\boldsymbol{\sigma}}_m$.
The present MT scheme is based on the additive interaction law first developed and detailed by \cite{Mercier09} {and recalled in section \ref{sec23}}. {The solution algorithm is summarized in  \ref{Sec:App-Alg}.}

The goal of the present work is to revisit this linearization scheme by making use of the second moments of stresses instead of only the first moments of stresses.

\subsection{New tangent linearization using second moments of stresses \label{Sec:Tangent}}
 \label{sec32}

Let us now evaluate how the MT scheme based on the additive interaction law can incorporate second moments of stresses within the viscoplastic tangent compliance and the back-extrapolated strain rate.

Following \cite{Suquet95}, \cite{Masson20} and \cite{Berbenni21}, the measure for the second moments of deviatoric stresses is the scalar invariant $S_m$: 
\begin{equation}\label{Eq:SecondMoment}
 \bar{\bar{\sigma}}_{eq}=\sqrt{3/2\mathbb{I}_D\cdot\overline{(\boldsymbol{\sigma}\otimes\boldsymbol{\sigma})}_m}=\sqrt{3/2 \overline{(\mathbf{s}\cdot\mathbf{s})}_m}=\sqrt{3/2 S_m}
 \end{equation}
In addition, we state that the viscous tangent and secant moduli of the matrix phase defined in Eq. \eqref{Eq:moduli} can be evaluated using this measure (i.e. $S_m$) of the second moments of stresses. Therefore, the tangent and secant moduli become 
\begin{equation}\label{Eq:moduli2}
2\mu^{tg}(\bar{\bar{\sigma}}_{eq})=\frac{2M\sigma_Y}{3\dot{\varepsilon}_0}
\left(\frac{\left<\,\bar{\bar{\sigma}}_{eq}-\sigma_Y\,\right>}{\sigma_Y}\right)^{1-\frac{1}{M}} ,\quad 2\mu^{sec}(\bar{\bar{\sigma}}_{eq})=\frac{2\bar{\bar{\sigma}}_{eq}}{3\dot{\varepsilon}_0}
\left(\frac{\sigma_Y}{\left<\,\bar{\bar{\sigma}}_{eq}-\sigma_Y\,\right>}\right)^{\frac{1}{M}},
\end{equation}
while the stress direction tensor ${\mathbf{N}}_s$ and deviatoric stress $\mathbf{s}$ are still calculated using the first moments of stress tensor in the matrix. Therefore the viscoplastic compliance and back extrapolated strain rate can be written in the condensed forms:
 \begin{equation}\label{Eq:Mtan2M}
 \mathbb{M}^{v-tg}=\mathbb{M}^{v-tg}(\bar{\mathbf{N}}_s,\bar{\bar{{\sigma}}}_{eq})
 \end{equation}
 \begin{equation}
 \label{Eq:epref2M}
    \dot{\boldsymbol{\varepsilon}}^{v-ref}=\dot{\boldsymbol{\varepsilon}}^{v-ref}(\bar{\bar{\sigma}}_{eq},\bar{\mathbf{s}}_m)=\left(\frac{1}{2\mu^{sec}(\bar{\bar{\sigma}}_{eq})}-\frac{1}{2\mu^{tg}(\bar{\bar{\sigma}}_{eq})}\right)\bar{\mathbf{s}}_m\,.
\end{equation}
Finally, the proposed linearization of the viscoplastic response becomes:
\begin{equation}\label{Eq:ViscoLinb}
\dot{\boldsymbol{\varepsilon}}^v=\mathbb{M}^{v-tg}(\bar{\mathbf{N}}_s,\bar{\bar{{\sigma}}}_{eq})
\cdot\boldsymbol{\sigma}+\dot{\boldsymbol{\varepsilon}}^{v-ref}(\bar{\bar{\sigma}}_{eq},\bar{\mathbf{s}}_m)
\end{equation} 
Averaging $\dot{\boldsymbol{\varepsilon}}^{v}$ given in Eq. \eqref{Eq:ViscoLinb} over the matrix phase, and using Eqs. \eqref{Eq:Mtan2M}-\eqref{Eq:epref2M}, the viscous matrix response (in an average sense) becomes: 
\begin{equation}\label{Eq:NortonPerzyna-2S}
{\dot{\bar{\boldsymbol{\varepsilon}}}^v}_m=\frac{3\dot{\varepsilon}_{0}}{2\bar{\bar{\sigma}}_{eq}}\left(\frac{\left<\,\bar{\bar{\sigma}}_{eq}-\sigma_Y\,\right>}{\sigma_Y}\right)^{\frac{1}{M}}\bar{\mathbf{s}}_m=\frac{1}{2\mu^{sec}(\bar{\bar{\sigma}}_{eq})}\bar{\mathbf{s}}_m
\end{equation}
  
In order to calculate the current value of the quantity $S_m$, {similarly to \cite{Masson20} and \cite{Berbenni21} in the case of modified secant formulations}, the Hill-Mandel's lemma is used in the form:
 \begin{equation}\label{Eq:Hill}
 \boldsymbol{\Sigma}\cdot\dot{\mathbf{E}}=\overline{\boldsymbol{\sigma}\cdot\dot{\boldsymbol{\varepsilon}}}=
 f_i\overline{(\boldsymbol{\sigma}\cdot\dot{\boldsymbol{\varepsilon}})}_i+f_m\overline{(\boldsymbol{\sigma}\cdot\dot{\boldsymbol{\varepsilon}})}_m=f_i\bar{\boldsymbol{\sigma}}_i\cdot\dot{\bar{\boldsymbol{\varepsilon}}}_i+f_m\overline{(\boldsymbol{\sigma}\cdot\dot{\boldsymbol{\varepsilon}})}_m
 \end{equation} 
As in \cite{Masson20} and \cite{Berbenni21}, we assume that for the inclusion phase, the first moments of stresses and strain are sufficient to evaluate the first term of the right hand side of Eq. \eqref{Eq:Hill}. {Such assumption seems to work} well for elastic inclusions {(which is the case for the particulate-reinforced composite materials considered in all examples in section \ref{Sec:Results})} or voids.  
Isotropic elasticity holds for both the matrix phase and inclusions. Therefore, the matrix contribution in Eq \eqref{Eq:Hill} expands as follows:
\begin{equation}\label{Eq:Hill-matrix}
\overline{(\boldsymbol{\sigma}\cdot\dot{\boldsymbol{\varepsilon}})}_m=\frac{1}{9 K_m^e}\overline{((\mathrm{tr}\boldsymbol{\sigma})(\mathrm{tr}\dot{\boldsymbol{\sigma}}))}_m+\frac{1}{2\mu_m^e}\overline{(\mathbf{s}\cdot\dot{\mathbf{s}})}_m+\overline{(\boldsymbol{\sigma}\cdot\dot{\boldsymbol{\varepsilon}}^v)}_m
 \end{equation}

To treat the first term, as in \cite{Masson20} and \cite{Berbenni21} we assume that there is no stress fluctuations in the hydrostatic part of the stress tensor for the matrix phase, so that $\overline{((\mathrm{tr}\boldsymbol{\sigma})(\mathrm{tr}\dot{\boldsymbol{\sigma}}))}_m\approx(\mathrm{tr}\bar{\boldsymbol{\sigma}}_m)(\mathrm{tr}\dot{\bar{\boldsymbol{\sigma}}}_m)$. The second moments of stresses are only present in the deviatoric part. In addition, to calculate the second term with stress fluctuations in the deviatoric part of the stress tensor, we use $S_m=\overline{(\mathbf{s}\cdot\mathbf{s})}_m$, such that $\overline{(\mathbf{s}\cdot\dot{\mathbf{s}})}_m=\dfrac{1}{2}\dot{S}_m=\dfrac{1}{2}\dot{\overline{(\mathbf{s}\cdot\mathbf{s})}_m}$. 

Let us now expand the third term $\overline{(\boldsymbol{\sigma}\cdot\dot{\boldsymbol{\varepsilon}}^v)}_m$ related to the viscoplastic behavior of the matrix. Substituting Eq \eqref{Eq:Affine} with specifications given by Eqs. \eqref{Eq:Mtan2M} and \eqref{Eq:epref2M}, we obtain:
\begin{equation}\label{Eq:evHill}
\overline{(\boldsymbol{\sigma}\cdot\dot{\boldsymbol{\varepsilon}}^v)}_m=
\frac{1}{2\mu^{tg}(\bar{\bar{\sigma}}_{eq})}\overline{(\mathbf{s}\cdot\bar{\mathbf{N}}_s)^2}_m+
\frac{1}{2\mu^{seq}(\bar{\bar{\sigma}}_{eq})}\left(\overline{(\mathbf{s}\cdot\mathbf{s})}_m-\overline{(\mathbf{s}\cdot\bar{\mathbf{N}}_s)^2}_m \right)+\left(\frac{1}{2\mu^{sec}(\bar{\bar{\sigma}}_{eq})}-\frac{1}{2\mu^{tg}(\bar{\bar{\sigma}}_{eq})}\right)\bar{\mathbf{s}}_m\cdot\bar{\mathbf{s}}_m
\end{equation}
After simple manipulations the relation can be rewritten in the following form:
\begin{equation}\label{Eq:evHillav}
\overline{(\boldsymbol{\sigma}\cdot\dot{\boldsymbol{\varepsilon}}^v)}_m=
\frac{1}{2\mu^{seq}(\bar{\bar{\sigma}}_{eq})} S_m+
\left(\frac{1}{2\mu^{tg}(\bar{\bar{\sigma}}_{eq})}-\frac{1}{2\mu^{seq}(\bar{\bar{\sigma}}_{eq})}\right)\left((\overline{{\mathbf{s}}\cdot\bar{\mathbf{N}}_s)^2}_m-\bar{\mathbf{s}}_m\cdot\bar{\mathbf{s}}_m\right)
\end{equation}
Let us note that for homogeneous stress field in the matrix, $(\overline{{\mathbf{s}}\cdot\bar{\mathbf{N}}_s)^2}_m=\bar{\mathbf{s}}_m\cdot\bar{\mathbf{s}}_m$. 
For heterogeneous stress field and for limited volume fraction of inclusions, which is the case considered in the present work, it can be reasonably expected that the second term of the right hand side of Eq. \eqref{Eq:evHillav} is negligible ($(\overline{{\mathbf{s}}\cdot\bar{\mathbf{N}}_s)^2}_m-\bar{\mathbf{s}}_m\cdot\bar{\mathbf{s}}_m\approx 0$) compared to the first term. Thus we assume that 
\begin{equation}\label{Eq:evterm-app}
\overline{(\boldsymbol{\sigma}\cdot\dot{\boldsymbol{\varepsilon}}^v)}_m\approx
\frac{1}{2\mu^{seq}(\bar{\bar{\sigma}}_{eq})} S_m.   
\end{equation}

{Using Eq. \eqref{Eq:Hill} together with Eqs. \eqref{Eq:Hill-matrix} and \eqref{Eq:evterm-app}, the first time derivative of $S_m$ is obtained in a simple form similar to the modified secant formulations of \cite{Masson20} and \cite{Berbenni21} (see  \ref{Sec:App-Sec} for a comparison):}
\begin{equation}\label{Eq:HillResult}
\frac{f_m}{4\mu_m^e}\dot{S}_m=\boldsymbol{\Sigma}\cdot\dot{\mathbf{E}}-f_i\bar{\boldsymbol{\sigma}}_i\cdot\dot{\bar{\boldsymbol{\varepsilon}}}_i- \frac{f_m}{9 K_m^e}(\mathrm{tr}\bar{\boldsymbol{\sigma}}_m)(\mathrm{tr}\dot{\bar{\boldsymbol{\sigma}}}_m)-f_m\frac{1}{2\mu^{sec}(\bar{\bar{\sigma}}_{eq})}S_m
\end{equation}

With Eq \eqref{Eq:HillResult}, the update of the second moment invariant $S_m$ could be performed according to the classic forward Euler scheme:
\begin{equation}\label{Eq:linupdate}
S_m(t+\Delta t)=S_m(t)+\dot{S}_m(t)\Delta t\,.
\end{equation}
However, it is well known that the explicit forward Euler scheme needs very small time increment $\Delta t$ to achieve good convergence, which is not computationally efficient.  Therefore, when implementing the model, a refined method of integration, involving the second time derivative of $S_m$, is used, which leads to the enhancement of computational efficiency of the solution procedure. The proposed method is reported in \ref{Sec:App-Alg}, where the influence of the time step on results for both integration methods is illustrated.

\subsection{Discussion} 
The present approach can be related somehow to \cite{PonteCastaneda2002} and its second-order homogenization method based on a variational approach. Let us recall the definition of the fourth order stress covariance tensor for the matrix phase:
\begin{equation}\label{Eq:stress-cov}
\mathbb{C}_{\sigma}^m=\overline{(\boldsymbol{\sigma}-\bar{\boldsymbol{\sigma}}_m)\otimes(\boldsymbol{\sigma}-\bar{\boldsymbol{\sigma}}_m)}_m= \overline{\boldsymbol{\sigma}\otimes\boldsymbol{\sigma}}_m-\bar{\boldsymbol{\sigma}}_m\otimes\bar{\boldsymbol{\sigma}}_m  
\end{equation}
and its scalar invariants:
\begin{eqnarray}\label{Eq:delta1}
\delta_{\rm{P}}^m&=&\mathbb{I}_{\rm{P}}\cdot\mathbb{C}_{\sigma}^m=\frac{1}{3}\left(\overline{(\rm{tr}\boldsymbol{\sigma})^2}_m-(\rm{tr}\boldsymbol{\bar{\sigma}}_m)^2\right)\\ \label{Eq:deltaD}
\delta_{\rm{D}}^m&=&\mathbb{I}_{\rm{D}}\cdot\mathbb{C}_{\sigma}^m={\overline{(\mathbf{s}\cdot\mathbf{s})}_m-\bar{\mathbf{s}}_m\cdot\bar{\mathbf{s}}_m}=\frac{2}{3}(\bar{\bar{\sigma}}_{eq}^2-\bar{\sigma}_{eq}^2)\\ \label{Eq:deltaIID}
\delta_{\rm{D}}^{|| m}&=&(\bar{\mathbf{N}}_s\otimes\bar{\mathbf{N}}_s)\cdot\mathbb{C}_{\sigma}^m=\overline{({\mathbf{s}}\cdot\bar{\mathbf{N}}_s)^2}_m-\bar{\mathbf{s}}_m\cdot\bar{\mathbf{s}}_m
\\ \label{Eq:delta4}
\delta_{\rm{D}}^{\perp m}&=&\delta_{\rm{D}}^m-\delta_{\rm{D}}^{|| m}=(\mathbb{I}_D-\bar{\mathbf{N}}_s\otimes\bar{\mathbf{N}}_s)\cdot\mathbb{C}_{\sigma}^m=\overline{\left(\mathbf{s}\cdot\mathbf{s}-({\mathbf{s}}\cdot\bar{\mathbf{N}}_s)^2\right)}_m
\end{eqnarray}
All the derivations to obtain the evolution law of $S_m$, as specified by Eq. \eqref{Eq:HillResult}, were possible assuming the following conditions for the invariants of the stress covariance tensor:
\begin{equation}\label{Eq:ass-cov}
\delta_{\rm{P}}^m=\delta_{\rm{D}}^{|| m}=0\,\quad\textrm{and}\quad \delta_{\rm{D}}^m=\delta_{\rm{D}}^{\perp m}\,.
\end{equation}
For the inclusion phase, the stress covariance tensor is assumed to be zero:
\begin{equation}
    \mathbb{C}_{\sigma}^i=\mathbb{O}\,.
\end{equation}
 Let us remark that the assumptions in Eqs. \eqref{Eq:ass-cov} impose ad-hoc restrictions on the stress fluctuation and need to be validated with respect to reference full-field solutions. Nevertheless, this allow us to avoid the minimization procedure inherent to the variational approach, see \cite{PonteCastaneda2002}. Therefore, the solution proceeds in a more direct way.

 {We also note that the present model has some similarities with the  modified affine theory developed for non linear composites by \cite{Brenner01}. In their model, \cite{Brenner01} also considered second order moments of stresses in the tangent compliance tensor. In contrast to the present proposal (see Eq. \eqref{Eq:Mtan2M}), the incorporation of the second moments is made in both the stress direction and in the equivalent stress (see page 652 therein). Moreover, they considered non linear elastic phases (a power law description), which could be atributed to a Norton’s type viscoplastic law without elasticity, in the framework of the self-consistent scheme for two-phase composites. The calculation of the second order moments of stresses (based on the strain energy) follow the 
equation (6) of their paper, which is different from our Eq. \eqref{Eq:HillResult} derived for two-phase particulate composites with elastic-viscoplastic behavior. Here, the study of the asymptotic regime (purely viscoplastic) using Eq. \eqref{Eq:HillResult} is obtained by setting 
$\dot{S}_m=0$ and $\rm{tr}\dot{\boldsymbol{\sigma}}_m=0$. Because of the different homogenization schemes (SC versus MT) and material 
behaviors (Norton's versus Perzyna's), quantitative comparisons between both approaches regarding this asymptotic regime are left for further studies.}

\section{Results\label{Sec:Results}}

In this section, the accuracy of the proposed modification of the additive tangent Mori-Tanaka scheme is compared with the original tangent formulation of  \cite{Mercier09} and \cite{Czarnota15}. {Numerical results obtained by full-field FFT analyses  available in \citep{Lahellec13,Masson20} are used as reference calculations. These authors performed calculations with a representative volume element containing 50 randomly distributed inclusions.} In all cases, a two-phase particulate composite is considered with the elastic-viscoplastic matrix described by the Perzyna-type law \eqref{Eq:Perzyna-seq} reinforced by purely elastic spherical inclusions. The volume fraction of inclusions is $17\%$. Material parameters of both phases are contained in Table \ref{Tab:mater-parameters}. Strain-controlled monotonic, cyclic and non-radial loading cases are analyzed. 

\begin{table}[h!]
	\begin{center}
		\begin{tabular}{c|lllll}
			&$K$ [GPa] & $\mu$ [GPa]  & ${\sigma}_Y$ [MPa] & $\dot{\varepsilon}_o$ [$\rm s^{-1}$] & $M$\\ \hline
			elastic-viscoplastic matrix &10 & 3 & 100 & 1 & $0.3$ \\ 
			elastic  inclusion &20 & 6 & --- & --- & ---\\
		\end{tabular}
		\caption{Material properties of both phases. The matrix phase is elastic-viscoplastic with the Perzyna-type law given by Eq. \eqref{Eq:Perzyna-seq} and inclusions are elastic. 
		}
		\label{Tab:mater-parameters}
	\end{center}
\end{table}

    \vspace{-0.5cm}
\subsection{Monotonic isochoric tension}

First, monotonic isochoric tension in direction $\mathbf{e}_3$ is analyzed. The process is specified by the overall strain given as:
\begin{equation}\label{Eq:isoch}
\mathbf{E}(t)=\frac{\dot{E} \cdot t}{2}(3\mathbf{e}_3\otimes\mathbf{e}_3-\mathbf{I})
\end{equation}
where $\dot{E}$ is a constant imposed strain rate. 

\begin{figure}[!h]
    \begin{tabular}{cc}
    (a) & (b)\\
        \includegraphics[angle=0,width=0.45\textwidth]{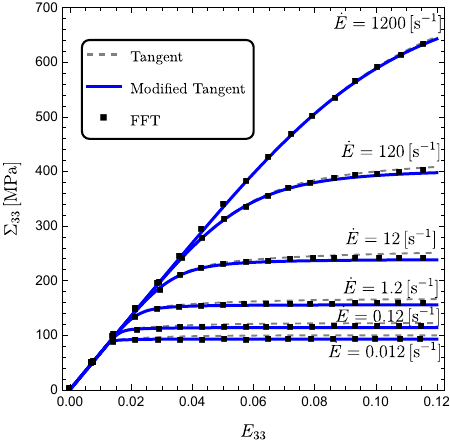}&\includegraphics[angle=0,width=0.45\textwidth]{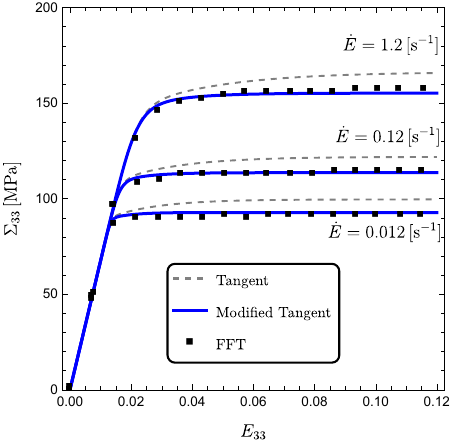}
    \end{tabular}
    \vspace{-0.5cm}
\caption{Macroscopic axial stress $\Sigma_{33}$ versus macroscopic axial strain $E_{33}$ during isochoric tension defined in Eq. \eqref{Eq:isoch}. Predictions obtained with the modified and original MT tangent additive schemes are presented for various strain rates. A two-phase composite with $17\%$ of elastic inclusion phase and an elastic-viscoplastic matrix of the Perzyna-type law \eqref{Eq:Perzyna} is considered. (a) macroscopic response for strain rate in the range 0.012 s$^{-1}$ to 1200 s$^{-1}$, b) for the lowest three strain rates. Reference full-field results obtained by FFT are taken from \cite{Lahellec13}. Material parameters of both phases are reported in Table \ref{Tab:mater-parameters}}\label{Fig:Monot}
\end{figure}

This macroscopic strain rate $\dot{E}$ is varied in the range between $0.012$ and $12$ $[\rm s^{-1}]$. The model predictions in terms of the overall axial stress $\Sigma_{33}$ vs. the overall axial strain $E_{33}$ are illustrated in Fig. \ref{Fig:Monot}. Predictions are displayed up to 0.12 total axial strain $E_{33}$.

 In Fig. \ref{Fig:Monot}, it is seen that independently of the strain rate, the stress estimates given by the modified tangent additive scheme (including the second moments of stresses) are softer than those predicted by the original scheme considering only the first moments of stresses. Moreover, as compared to the reference FFT results, the new formulation outperforms the original one especially at the lowest strain rates, which is clearly observed in Fig. \ref{Fig:Monot}b.

\subsection{Cyclic isochoric tension-compression}

The second example concerns the cyclic isochoric tension-compression test, where the overall strain path is still specified by Eq. \eqref{Eq:isoch}, while the strain rate $\dot{E}$ changes its sign, namely
\begin{equation}
    \dot{E}(t)=\left\{ \begin{array}{lcl}
    12\times 10^{-3}\, {\rm{s}^{-1}}&\textrm{if}& t\leq 10\, [{\rm{s}}]\,\,\textrm{or}\,\, t>30\, [{\rm{s}}]\\
    -12\times 10^{-3}\, {\rm{s}^{-1}}&\textrm{if}& 10\, [{\rm{s}}]<t\leq 30\, [{\rm{s}}]
    \end{array}
    \right. 
\end{equation}
The process is conducted up to time $t=40\,[{\rm{s}}]$.

\begin{figure}[!h]
\centering
    \includegraphics[angle=0,width=0.6\textwidth]{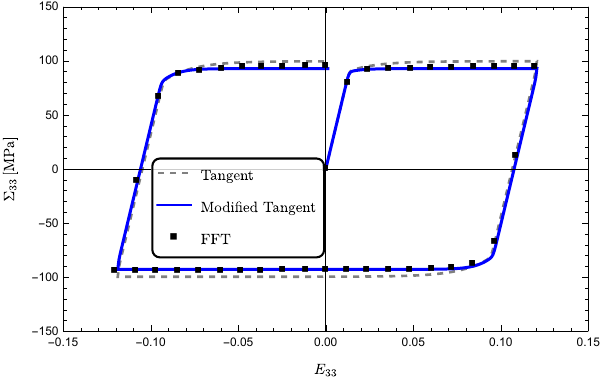}
\caption{Macroscopic axial stress $\Sigma_{33}$ vs. macroscopic axial strain $E_{33}$ in isochoric tension-compression cycle for a two-phase composite with $17\%$ of inclusion phase. Predictions of the modified and original additive tangent MT schemes are compared with the FFT results of \cite{Lahellec13}. The inclusion has an elastic behavior while the matrix has an elastic-viscoplastic response of the Perzyna-type, see Eq. \eqref{Eq:Perzyna}. Material parameters of both phases are reported in Table \ref{Tab:mater-parameters}}\label{Fig:cyclicO}
\end{figure}  

\begin{figure}[!h]
\centering
    \begin{tabular}{c}
        (a)\\
          \includegraphics[angle=0,width=0.6\textwidth]{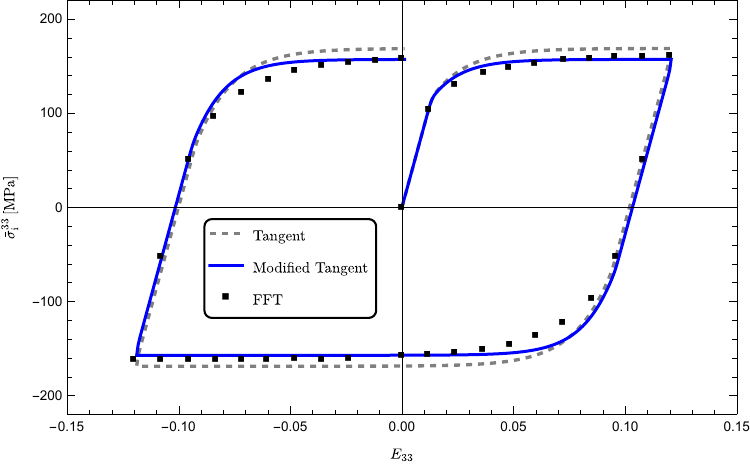}\\
          (b)\\
          \includegraphics[angle=0,width=0.6\textwidth]{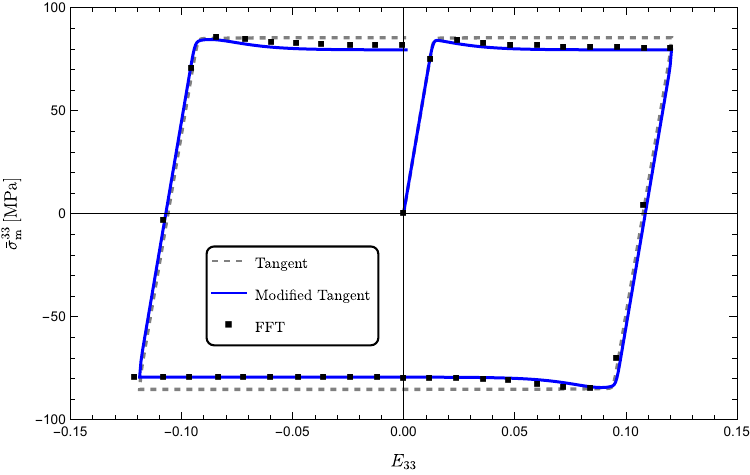}
    \end{tabular}
\caption{Mean axial stress $\bar{\sigma}_{33}$ (a) in the inclusion and (b) the matrix vs. macroscopic axial strain $E_{33}$ in isochoric tension-compression cycle for a two-phase composite with $17\%$ of inclusion phase. Predictions of the modified and original additive tangent MT schemes are compared with the FFT results of \cite{Lahellec13}. The inclusion has an elastic behavior while the matrix has an elastic-viscoplastic response of the Perzyna-type, see Eq. \eqref{Eq:Perzyna}. Material parameters of both phases are reported in Table \ref{Tab:mater-parameters}}\label{Fig:cyclic-m}
\end{figure}

\begin{figure}[!h]
  \centering  
        \includegraphics[angle=0,width=0.6\textwidth]{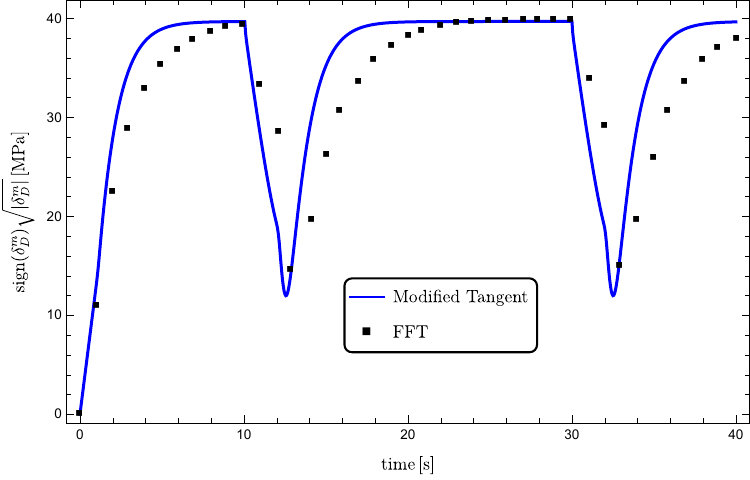}
\caption{Stress fluctuation in the matrix $\textrm{sign}(\delta^m_{\rm{D}})\sqrt{|\delta^m_{\rm{D}}|}$ (Eq \eqref{Eq:deltaD}) vs. time in isochoric tension-compression cycle  for a
two-phase composite with $17\%$ of inclusion phase. Predictions of the modified additive tangent MT schemes are compared with the FFT results of \cite{Lahellec13}. The inclusion has an elastic behavior while the matrix has an elastic-viscoplastic response of the Perzyna-type, see Eq. \eqref{Eq:Perzyna}. Material parameters of both phases are reported in Table \ref{Tab:mater-parameters}}\label{Fig:cyclic-f}
\end{figure} 

Fig. \ref{Fig:cyclicO} presents the evolution of the overall axial stress $\Sigma_{33}$ in terms of the overall axial strain $E_{33}$ during one cycle, as predicted by the original and modified additive tangent Mori-Tanaka models. The figure also includes the reference results obtained by the full-field FFT analyses reported by \cite{Lahellec13}. A perfect agreement between the modified tangent model and FFT results is seen, while the standard additive MT scheme slightly overpredicts the stress level.

In Fig. \ref{Fig:cyclic-m}, similar comparison is demonstrated regarding the estimates of the mean stress variation in both phases. It is seen that for both phases the new variant with modified tangent compliance in the additive scheme correctly predicts the stress level. In particular, for the matrix phase a slight softening after reaching the yield point, as observed in the reference FFT results, is correctly reproduced, contrary to the constant stress of the original first-moment formulation. The largest discrepancy between the mean-field models and the reference results is seen for the mean stress in the inclusion in the transient regime, when the viscous flow is initiated within the matrix phase. Nevertheless, it is clear that the predictive capabilities of the stress in the inclusion phase are improved during steady state regime of viscous plastic flow in the matrix, as compared to the approaches proposed by \cite{Lahellec13} and \cite{Berbenni21} (see also \ref{Sec:App-Sec}, Fig. \ref{Fig:InFlu-Sec}a). 
Let us remark that the predictions of the mean strain per phase are not discussed since, unfortunately, the respective reference results are not available in \cite{Lahellec13}.

Finally, to explain these slight differences, Fig. \ref{Fig:cyclic-f} shows the stress fluctuations over the matrix phase as a function of time. Following \cite{Lahellec13}, \cite{Masson20} and \cite{Berbenni21}, the time evolution of the square root of the deviatoric stress covariance invariant: $\textrm{sign}(\delta^m_{\rm{D}})\sqrt{|\delta^m_{\rm{D}}|}$ (let us note that $\delta^m_{\rm{D}}$ calculated from the mean-field model is not necessarily positive, while it is for the full-field reference calculations) serves as indicator for the fluctuations. A very good quantitative agreement between the modified additive MT scheme and the reference FFT results is observed, especially during the regimes with advanced viscoplastic flow in the matrix. Discrepancies are mainly visible in the transient regime, which correlates with discrepancies observed in estimates of the axial mean stress in the inclusion phase, see Fig. \ref{Fig:cyclic-m}a. Let us note that the evolution of $\delta^m_{\rm{D}}$ is not accessible with the original tangent formulation with first moments of stresses. When comparing with the literature, it can be verified that the observed differences are much smaller than for other approaches which provide access to stress fluctuations. For the same example, the rate-variational procedure (RVP) of \cite{Lahellec13} (see Fig. 10 therein) and the modified secant model of \cite{Berbenni21} (see Fig. 5 therein and \ref{Sec:App-Sec}, Fig. \ref{Fig:InFlu-Sec}b) both overestimate the value of the stress fluctuations, while the modified secant model of \cite{Masson20} (see Fig. 6 therein) visibly underpredicts the value of the stress fluctuations, with a small overshoot during the first loading due to an instability of this model at the beginning of the viscoplastic flow (according to these authors).

\subsection{Non-proportional loading}
Following \cite{Masson20}, in the last example the non-radial strain-controlled loading is considered in which shear is super-imposed on the isochoric axial loading:
\begin{equation}\label{Eq:non-radial}
\mathbf{E}(t)=\frac{\dot{E}_{33}(t)t}{2}(3\mathbf{e}_3\otimes\mathbf{e}_3-\mathbf{I})+\dot{E}_{13}(t) t(\mathbf
{e}_1\otimes\mathbf{e}_3+\mathbf
{e}_3\otimes\mathbf{e}_1+\mathbf
{e}_2\otimes\mathbf{e}_3+\mathbf
{e}_3\otimes\mathbf{e}_2)
\end{equation}
where
\begin{equation}
\dot{E}_{33}(t)=\left\{ \begin{array}{rcl}
    12\times 10^{-3}\, [{\rm{s}^{1-}}]&\textrm{if}& t\leq 10\, [{\rm{s}}]\\
    0 &\textrm{if}& 10 < t\leq 20\, [{\rm{s}}] \,\,\textrm{or}\,\, 30 < t\leq 40\, [{\rm{s}}]\\
    -12\times 10^{-3}\,  [{\rm{s}^{-1}}]&\textrm{if}& 20 < t\leq 30\, [{\rm{s}}]
    \end{array}
    \right.     
\end{equation}
and 
\begin{equation}
\dot{E}_{13}(t)=\left\{ \begin{array}{rcl}
    0\, &\textrm{if}& t\leq 10\,[s] \,\,\textrm{or}\,\,  20 < t\leq 30\, [{\rm{s}}] \\
    7.5\times 10^{-3}\,  [{\rm{s}^{-1}}] &\textrm{if}& 10 < t\leq 20\, [{\rm{s}}]  \\
    -7.5\times 10^{-3}\,  [{\rm{s}^{-1}}]&\textrm{if}& 30 < t\leq 40\, [{\rm{s}}]
    \end{array}
    \right.     
\end{equation}
This strain-controlled process was also analyzed by \cite{Lahellec13}, however, in the context of the rate-independent elastic-plastic matrix material. 

\begin{figure}[!h]
\centering
  \includegraphics[angle=0,width=0.6\textwidth]{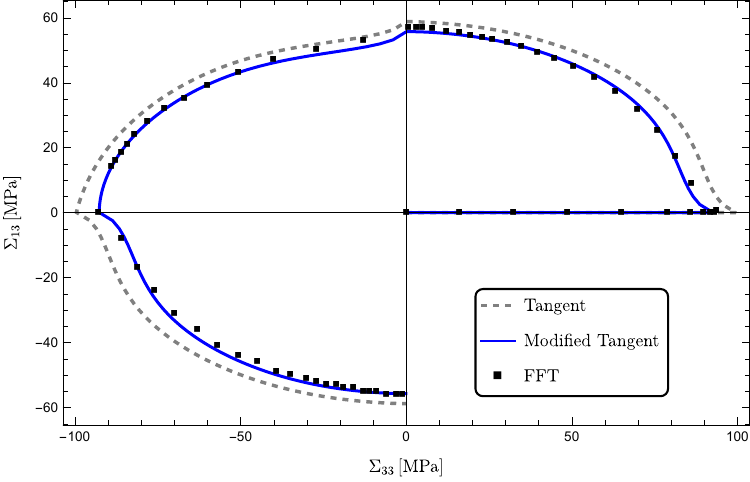}
\caption{Overall stress component $\Sigma_{13}$ as a function of $\Sigma_{33}$ for a non-proportional loading given in Eq. \eqref{Eq:non-radial}. Predictions of the modified and original additive tangent MT schemes are compared with the FFT results of \cite{Masson20}. The inclusion has an elastic behavior while the matrix has an elastic-viscoplastic response of the Perzyna-type, see Eq. \eqref{Eq:Perzyna}. Material parameters of both phases are reported in Table \ref{Tab:mater-parameters}}\label{Fig:non-radial}
\end{figure}

Figure \ref{Fig:non-radial} presents the overall stress evolution in the plane $(\Sigma_{33},\Sigma_{13})$. Similarly to previous examples, the modified tangent MT additive model delivers an improved prediction with respect to the standard first-moment formulation as compared to the reference full-field results.

\section{Conclusions}

In the presented study, the incorporation of the second moments of stresses has been introduced for the first time into the tangent formulation of the additive Mori-Tanaka model for two-phase elastic-viscoplastic material. In contrast with the modified secant linearization, which is associated with an isotropic viscoplastic compliance, the modified tangent linearization of the viscoplastic laws has been performed starting from the anisotropic viscoplastic compliance as for the original tangent model. Then, the Hill-Mandel's lemma has been used to track the evolution of the second moments of stresses. To study the model performance, the results have been compared to the full-field numerical calculations and predictions of other models available in the literature. A very good performance of the modified tangent scheme based on second moments of stresses has been demonstrated from these benchmarks. It has been shown that it outperforms the original tangent formulation with first moments of stresses. Additionally, it also improves the results obtained from the methods based on a modified secant formulation.

{The present additive model with second order stress moments can be easily implemented in a FE software and directly in the time domain, as in \cite{Msolli16} and \cite{Sadowski21}, who used the tangent additive Mori-Tanaka law for elastic-viscoplastic composites based on a classical tangent linearization. As the main goal of the paper is to validate our proposition, the use of the proposed method for applications will be considered in a future contribution.}

\section*{Acknowledgements}
The research of K. Kowalczyk-Gajewska was partially supported by the {project} 2021/41/B/ST8/03345 of the National Science Centre, Poland. S. Berbenni and S. Mercier acknowledge the support of the National Research Agency (ANR), France, under the project ANOHONA (ANR-23-CE51-0047). 

\appendix

\section{Solution algorithms}\label{Sec:App-Alg}

\paragraph{Solution algorithm for the first moment tangent formulation} {Solution is found iteratively as proposed in \cite{Mercier09}. We assume that at time $t$, the macroscopic strain rate $\dot{\mathbf{E}}(t)$ and the current values of mean stresses in the phases $\bar{\boldsymbol{\sigma}}_k(t)$, $(k=i,m)$ are known.} {With this input data, at each time increment the following steps are performed:
\begin{enumerate}
\item[\textbf{Step 1:}] Using $\bar{\boldsymbol{\sigma}}_m(t)$ the fourth order viscous compliance tensor for the matrix phase, $\mathbb{M}^{v-tg}(\bar{\mathbf{N}}_s(t),\bar{{\sigma}}_{eq}(t))$ is calculated by Eq. \eqref{Eq:Mvtg}, 
\item[\textbf{Step 2:}] Using $\mathbb{M}^{v-tg}(\bar{\mathbf{N}}_s(t),\bar{{\sigma}}_{eq}(t))$ the viscous inverse Hill tensor $\mathbb{M}_{v}^*(\bar{\boldsymbol{\sigma}}_m(t))$ is specified by Eq. \eqref{Eq:MHilldef}. Let us note that because the viscous tangent compliance tensor $\mathbb{M}^{v-tg}$ is anisotropic, numerical integration needs to be performed to find  $\mathbb{P}_{v}^*$.
\item[\textbf{Step 3:}] Four quantities: $\dot{\bar{\boldsymbol{\varepsilon}}}_k(t)$, $\dot{\bar{\boldsymbol{\sigma}}}_k(t)$ ($k=i,m$) are obtained by solving the set of four equations ($f_k$ is the volume fraction of the phase $k$ $(k=i,m)$):
\begin{equation}\label{Eq:A1}
\dot{\bar{\boldsymbol{\varepsilon}}}_k=\mathbb{M}^e\cdot\dot{\bar{\boldsymbol{\sigma}}}_k+\mathbf{g}(\bar{\boldsymbol{\sigma}}_k)\,,\quad \dot{\mathbf{E}}=\sum_k f_k \dot{\bar{\boldsymbol{\varepsilon}}}_k
\end{equation}
\begin{equation}\label{Eq:A2}
 \dot{\bar{\boldsymbol{\varepsilon}}}_i-\dot{\bar{\boldsymbol{\varepsilon}}}_m=-\mathbb{M}_e^*\cdot(\dot{\bar{\boldsymbol{\sigma}}}_i-	\dot{\bar{\boldsymbol{\sigma}}}_m)-\mathbb{M}_v^*(\bar{\boldsymbol{\sigma}}_m)\cdot(\bar{\boldsymbol{\sigma}}_i-	\bar{\boldsymbol{\sigma}}_m)
\end{equation}
Indeed Eqs. \eqref{Eq:A1} and \eqref{Eq:A2} define a linear system for the strain and stress rates.
\item[Step 4:] An explicit update of the stress in each phase is adopted:
\begin{equation}
\bar{\boldsymbol{\sigma}}_k(t+\Delta t)=\bar{\boldsymbol{\sigma}}_k(t)+\dot{\bar{\boldsymbol{\sigma}}}_k(t)\Delta t
\end{equation}
The local strains are updated similarly.
\item[Step 5:] The macroscopic stress is found by volume averaging over the phases:
\begin{equation}
 \boldsymbol{\Sigma}(t+\Delta t)=\sum_k f_k \bar{\boldsymbol{\sigma}}_k(t +\Delta t)\,.   
\end{equation}
\end{enumerate}}

\paragraph{Solution algorithm for the tangent formulation with the second moment of stress} Similarly to first moment formulation, the solution proceeds iteratively. Accordingly, we assume that at time $t$, the macroscopic strain rate $\dot{\mathbf{E}}(t)$, the current values of mean stresses in the phases $\bar{\boldsymbol{\sigma}}_k(t)$, $(k=i,m)$ and the scalar invariant of second moment of stress $S_m(t)$ are known. As for the classical model, the four quantities $\dot{\bar{\boldsymbol{\varepsilon}}}_k(t)$, $\dot{\bar{\boldsymbol{\sigma}}}_k(t)$ ($(k=i,m)$) are obtained by solving the set of four equations ($f_k$ is the volume fraction of the phase $k$):
\begin{equation}\label{Eq:2Mset1}
\dot{\bar{\boldsymbol{\varepsilon}}}_m=\mathbb{M}^e_m\cdot\dot{\bar{\boldsymbol{\sigma}}}_m+\overline{\dot{\boldsymbol{\varepsilon}}^v}_m(\bar{\bar{{\sigma}}}_{eq},\bar{\mathbf{s}}_m) 
\,,\quad
\dot{\bar{\boldsymbol{\varepsilon}}}_i=\mathbb{M}^e_i\cdot\dot{\bar{\boldsymbol{\sigma}}}_i+\mathbf{g}(\bar{\boldsymbol{\sigma}}_i)\,,\quad \dot{\mathbf{E}}=\sum_k f_k \dot{\bar{\boldsymbol{\varepsilon}}}_k
\end{equation}
\begin{equation}\label{Eq:2Mset2}
\dot{\bar{\boldsymbol{\varepsilon}}}_i-\dot{\bar{\boldsymbol{\varepsilon}}}_m=-\mathbb{M}_e^*\cdot(\dot{\bar{\boldsymbol{\sigma}}}_i-	\dot{\bar{\boldsymbol{\sigma}}}_m)-\mathbb{M}_v^*(\bar{\mathbf{N}}_s,\bar{\bar{\sigma}}_{eq})\cdot(\bar{\boldsymbol{\sigma}}_i-	\bar{\boldsymbol{\sigma}}_m)
\end{equation}
Similarly to the first-moment formulation, an explicit update of the stress and strain in each phase is adopted:
\begin{equation}\label{Eq:update-sig}
\bar{\boldsymbol{\sigma}}_k(t+\Delta t)=\bar{\boldsymbol{\sigma}}_k(t)+\dot{\bar{\boldsymbol{\sigma}}}_k(t)\Delta t\,,\quad\bar{\boldsymbol{\varepsilon}}_k(t+\Delta t)=\bar{\boldsymbol{\varepsilon}}_k(t)+\dot{\bar{\boldsymbol{\varepsilon}}}_k(t)\Delta t 
\end{equation}
The macroscopic stress can then be found as:
\begin{equation}\label{Eq:up-OverStress}
\boldsymbol{\Sigma}(t+\Delta t)=\sum_k f_k \bar{\boldsymbol{\sigma}}_k(t+\Delta t).
\end{equation}
{However, contrary to the formulation based solely on the first moments, the updated value of $S_m$ is also needed. As mentioned in Section \ref{Sec:Tangent}, it is observed that a simple linear update of $S_m$ as specified by Eq. \eqref{Eq:linupdate}, in which $\dot{S}_m(t)$ is given by the differential equation \eqref{Eq:HillResult}, needs very small time increment $\Delta t$ to achieve good convergence of the results. Therefore, when implementing the model, a refined method of integration is used which leads to the enhancement of computational efficiency.}

\begin{figure}[!b]
\centering
(a)\\
\includegraphics[angle=0,width=0.6
\textwidth]{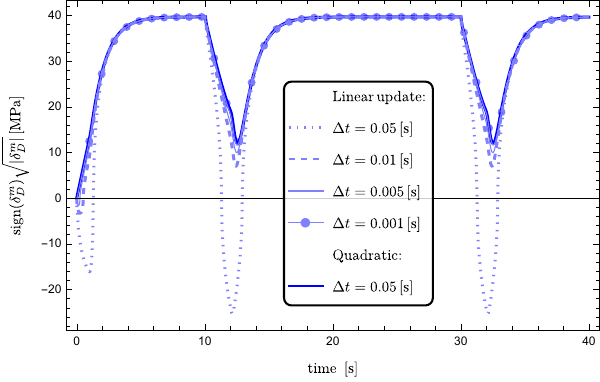}\\
    \begin{tabular}{cc}    
    (b)&(c)\\
    \includegraphics[angle=0,width=0.3\textwidth]{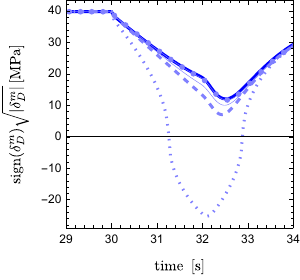}&
        \includegraphics[angle=0,width=0.3\textwidth]{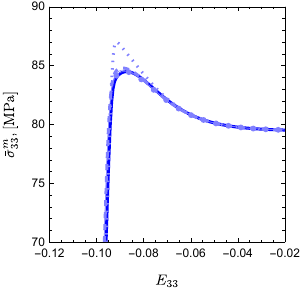}      
  \end{tabular}
\caption{\footnotesize{ (a) Evolution of stress fluctuations in the matrix: $\textrm{sign}(\delta^m_{\rm{D}})\sqrt{|\delta^m_{\rm{D}}|}$ (Eq. \eqref{Eq:deltaD}) over time integrated using linear update of $S_m$ (Eq. \eqref{Eq:linupdate}) for different time increment $\Delta t$ or using the quadratic update Eq. \eqref{Eq:Sm-update}. Results are obtained by the modified additive tangent MT scheme. Isochoric tension-compression cycle for a two-phase composite (see Section \ref{Sec:Results}). (b) zoom on the stress fluctuations during tensile re-loading regime and (c) corresponding curve of the mean matrix stress vs. macroscopic strain. Results for the quadratic update for all time steps: $0.05,0.01,0.005$ and $0.001$ [s] overlap with each other.}\label{Fig:algorithm}}
\end{figure}

For this purpose it is proposed to include the quadratic term in the respective Taylor expansion of $S_m$ and use the following update formula:
 \begin{equation}\label{Eq:Sm-update}
 S_m(t+\Delta t)=S_m(t)+\dot{S}_m(t)\Delta t+\frac{1}{2}\ddot{S}_m(t) \Delta t^2\,.
 \end{equation}
{where we assume that, although a linear update of ${\bar{\boldsymbol{\sigma}}}_m$ \eqref{Eq:update-sig} is sufficient, the quadratic correction is necessary for $S_m$, mainly because it is a quadratic function of $\mathbf{s}$. To update $S_m$ with Eq. \eqref{Eq:Sm-update}, one needs to evaluate its second derivative. To do so, the Hill-Mandel's lemma is used in its fully rate form, with similar approximation as before concerning the inclusion response, namely} 
\begin{equation}\label{Eq:Hill2}
 \dot{\boldsymbol{\Sigma}}\cdot\dot{\mathbf{E}}=\overline{\dot{\boldsymbol{\sigma}}\cdot\dot{\boldsymbol{\varepsilon}}}=
 f_i\overline{(\dot{\boldsymbol{\sigma}}\cdot\dot{\boldsymbol{\varepsilon}})}_i+f_m\overline{(\dot{\boldsymbol{\sigma}}\cdot\dot{\boldsymbol{\varepsilon}})}_m=f_i\dot{\bar{\boldsymbol{\sigma}}}_i\cdot\dot{\bar{\boldsymbol{\varepsilon}}}_i+f_m\overline{(\dot{\boldsymbol{\sigma}}\cdot\dot{\boldsymbol{\varepsilon}})}_m
 \end{equation} 
where the matrix contribution is calculated as:
\begin{equation}
\overline{(\dot{\boldsymbol{\sigma}}\cdot\dot{\boldsymbol{\varepsilon}})}_m=\frac{1}{9 K_m^e}\overline{((\mathrm{tr}\dot{\boldsymbol{\sigma}})^2)}_m+\frac{1}{2\mu_m^e}\overline{(\dot{{\mathbf{s}}}\cdot\dot{\mathbf{s}})}_m+\overline{(\boldsymbol{\dot{\sigma}}\cdot\dot{\boldsymbol{\varepsilon}}^v)}_m
 \end{equation}
together with:
\begin{equation}\label{Eq:evHill2}
\overline{(\boldsymbol{\dot{\sigma}}\cdot\dot{\boldsymbol{\varepsilon}}^v)}_m=\frac{1}{2\mu^{seq}(\bar{\bar{\sigma}}_{eq})}\overline{(\dot{\mathbf{s}}\cdot\mathbf{s})}_m+
\left(\frac{1}{2\mu^{tg}(\bar{\bar{\sigma}}_{eq})}-\frac{1}{2\mu^{sec}(\bar{\bar{\sigma}}_{eq})}\right)\left((\overline{\dot{\mathbf{s}}\cdot\bar{\mathbf{N}}_s)({\mathbf{s}}\cdot\bar{\mathbf{N}}_s)}_m-\dot{\bar{\mathbf{s}}}_m\cdot\bar{\mathbf{s}}_m\right)\,.
\end{equation} 
Similarly as for Eq. \eqref{Eq:evHillav}, observing that for the homogeneous stress field: $(\overline{\dot{\mathbf{s}}\cdot\bar{\mathbf{N}}_s)({\mathbf{s}}\cdot\bar{\mathbf{N}}_s)}_m=\dot{\bar{\mathbf{s}}}_m\cdot\bar{\mathbf{s}}_m$, we assume that the second term in the above formula is negligible as compared to the first one: $(\overline{\dot{\mathbf{s}}\cdot\mathbf{N}_s)({\mathbf{s}}\cdot\mathbf{N}_s)}_m-\dot{\bar{\mathbf{s}}}_m\cdot\bar{\mathbf{s}}_m\approx 0$\footnote{Let us note that this approximation is a direct consequence of assumption \eqref{Eq:ass-cov} if $\dot{\bar{\mathbf{N}}}_s=0$.}. 
{On the other hand, calculating $\ddot{S}_m$, we find:
\begin{equation}
    \ddot{S}_m=2\overline{\left(\mathbf{s}\cdot\ddot{\mathbf{s}}+\dot{\mathbf{s}}\cdot\dot{\mathbf{s}}\right)}_{m}= 2\overline{\left(\mathbf{s}\cdot\ddot{\mathbf{s}}\right)}_m+ 2\overline{\left(\dot{\mathbf{s}}\cdot\dot{\mathbf{s}}\right)}_m
\end{equation}
and assume that the first term is negligible as compared with the second one so that $\ddot{S}_m\approx 2\overline{\left(\dot{\mathbf{s}}\cdot\dot{\mathbf{s}}\right)}_m$. This latter assumption is in line with the proposed update \eqref{Eq:update-sig} of mean stress $\bar{\boldsymbol{\sigma}}_m$ in which the quadratic term is neglected.}

{Next, following} the {same approximation} concerning the hydrostatic part of stress within the matrix as in {Eq \eqref{Eq:HillResult}} and using {Eqs. \eqref{Eq:Hill2}} and \eqref{Eq:evHill2} with the second term neglected, the second derivative of $S_m$ is obtained:
\begin{equation}\label{Eq:HillResult2}
\frac{f_m}{4\mu_m^e}\ddot{S}_m=\dot{\boldsymbol{\Sigma}}\cdot\dot{\mathbf{E}}-f_i\dot{\bar{\boldsymbol{\sigma}}}_i\cdot\dot{\bar{\boldsymbol{\varepsilon}}}_i- \frac{f_m}{9 K_m^e}(\mathrm{tr}\dot{\bar{\boldsymbol{\sigma}}}_m)^2-f_m\frac{1}{4\mu^{sec}(\bar{\bar{\sigma}}_{eq})}\dot{S}_m\,,
\end{equation}
{which together with Eq. \eqref{Eq:HillResult} can be used in Eq. \eqref{Eq:Sm-update} to update $S_m$.}

The efficiency of the proposed enhancement of the solution algorithm is demonstrated in Fig. \ref{Fig:algorithm}. Fig. \ref{Fig:algorithm}~a,b shows that results are highly dependent on the time step when a linear update of $S_m$ (Eq. \ref{Eq:linupdate}) is accounted for. Therefore, for large time step one observes significant errors providing negative instead of positive values of $\delta^m_{\rm{D}}$. This affects the correct estimation of the yielding point in the matrix phase as seen in Fig. \ref{Fig:algorithm}~c. It should be mentioned that for the considered example, in the case of the quadratic update of $S_m$ \eqref{Eq:Sm-update} the reduction of the time step from $\Delta t=0.05$ to $0.001$ [s] did not change the predictions in the observable way. Obviously, for sufficiently small time steps,  predictions obtained by linear \eqref{Eq:linupdate} or quadratic \eqref{Eq:Sm-update} update merge to each other. 

{The solution algorithm for the modified tangent scheme can be summarized as follows. We assume that at time $t$ the required input data i.e. the macroscopic strain rate $\dot{\mathbf{E}}(t)$, the current values of mean stresses in the phases $\bar{\boldsymbol{\sigma}}_k(t)$, $(k=i,m)$ and the second moment invariant $S_m(t)$ are known. At each time increment the following steps are performed:
\begin{enumerate}
\item[\textbf{Step 1:}] Using $\bar{\boldsymbol{\sigma}}_m(t)$ and $S_m(t)$ the viscous compliance tensor of the matrix $\mathbb{M}^{v-tg}(\bar{\mathbf{N}}_s(t),\bar{\bar{{\sigma}}}_{eq}(t))$ is calculated, 
\item[\textbf{Step 2:}] Using $\mathbb{M}^{v-tg}(\bar{\mathbf{N}}_s(t),\bar{\bar{{\sigma}}}_{eq}(t))$ the viscous inverse Hill tensor $\mathbb{M}_{v}^*$ is numerically evaluated by Eq. \eqref{Eq:MHilldef} and \eqref{Eq:PdefAni}. 
\item[\textbf{Step 3:}] Four quantities: $\dot{\bar{\boldsymbol{\varepsilon}}}_k(t)$, $\dot{\bar{\boldsymbol{\sigma}}}_k(t)$ ($k=i,m$) are obtained by solving the linear system of four equations \eqref{Eq:2Mset1}-\ref{Eq:2Mset2}.
\item[\textbf{Step 4:}] An explicit update \eqref{Eq:update-sig} of the stress and strain in each phase is adopted.
\item[\textbf{Step 5:}] The macroscopic stress $\boldsymbol{\Sigma}(t+\Delta t)$ is found by volume averaging over the phases as specified by  Eq. \eqref{Eq:up-OverStress}.
\item[\textbf{Step 6:}] The first and second time derivatives: $\dot{S}_m(t)$ and  $\ddot{S}_m(t)$, are calculated using subsequently Eq. \eqref{Eq:HillResult} and Eq. \eqref{Eq:HillResult2}.
\item[\textbf{Step 7:}] An explicit update of $S_m(t+\Delta t)$ is performed according to the enhanced update formula \eqref{Eq:Sm-update}.
\end{enumerate}}

\section{The additive modified secant Mori-Tanaka model and comparison with the modified secant model of \cite{Berbenni21}}\label{Sec:App-Sec}

\paragraph{Secant linearization using first moments} It is observed that, instead of the tangent linearization, the viscous law \eqref{Eq:Perzyna} can be rewritten in a linearized form in a secant way, namely
\begin{equation}\label{Eq:secant}
\dot{\boldsymbol{\varepsilon}}^v=\mathbb{M}^{v-sec}(\boldsymbol{\sigma})\cdot\boldsymbol{\sigma}\,.
\end{equation} 
From the constitutive viscoplastic law \eqref{Eq:Perzyna} we can define an \emph{isotropic} viscoplastic secant compliance tensor  as follows
\begin{equation}
\label{Eq:Mvsec} \mathbb{M}^{v-sec}(\boldsymbol{\sigma})= \mathbb{M}^{v-sec}({\sigma}_{eq})=\frac{3\dot{\varepsilon}_{eq}}{2\sigma_{eq}}\mathbb{I}_D=\frac{1}{2\mu^{sec}(\sigma_{eq})}
\mathbb{I}_D
\end{equation}
where the relation $\mu^{sec}(\sigma_{eq})$ is given by Eq. \eqref{Eq:moduli}$_2$.
Let us note that the definition above justifies the notation applied earlier in Eq. \eqref{Eq:moduli}.

Now, to apply the solution of the Eshelby inclusion problem, within the classical formulation based on the first moments (mean values), we state that the viscous secant compliance $\mathbb{M}^{v-sec}(\sigma_{eq})$ for the matrix phase is approximated by quantities calculated using the mean stress in the matrix, namely:
\begin{equation}
\dot{\boldsymbol{\varepsilon}}^v=\mathbb{M}^{v-sec}({\bar{\sigma}}_{eq})\cdot\boldsymbol{\sigma}=\frac{1}{2\mu^{\rm{sec}}(\bar{\sigma}_{eq})}\mathbf{s}
\end{equation} 
where $\bar{\sigma}_{eq}$ is defined by Eq. \eqref{Eq:tan-first}.
Also in this case, either form of the viscoplastic constitutive relation: \eqref{Eq:Perzyna} or \eqref{Eq:NortonPerzyna-N} is equivalent with \eqref{Eq:secant} when written for mean values. 

Solution algorithm for the first-moment secant formulation is identical to the one used for the tangent formulation based on the first moments with an important advantage of the former from the point of view of computational efficiency. Namely, the numerical integration to obtain the viscoplastic Hill tensor is not necessary. The closed form relation \eqref{Eq:MHill} can be used in this case as the secant compliance \eqref{Eq:Mvsec} is isotropic ($K\rightarrow \infty$ is applied in Eq. \eqref{Eq:MHill} to calculate $\mathbb{M}^*_v$).

\paragraph{Secant linearization using second moments (the modified secant model)} 
We may also apply the Eshelby result calculating viscoplastic secant linearization based on the second moment of stress. So, specifically, we state that the viscous secant compliance $\mathbb{M}^{v-sec}({\sigma}_{eq})$ is approximated as follows
\begin{equation}\label{Eq:Mseq2M}
\mathbb{M}^{v-sec}({\sigma}_{eq})\approx\mathbb{M}^{v-sec}(\bar{\bar{{\sigma}}}_{eq})=\frac{1}{2\mu^{\rm{sec}}(\bar{\bar{\sigma}}_{eq})}\mathbb{I}_D
\end{equation}
where $\bar{\bar{\sigma}}_{eq}$ is defined by Eq. \eqref{Eq:SecondMoment}. Thus the linearized form of relation \eqref{Eq:Perzyna} for the matrix material is
\begin{equation}\label{Eq:NortonPerzyna-2SS}
\dot{\boldsymbol{\varepsilon}}^v=\frac{1}{2\mu^{\rm{sec}}(\bar{\bar{\sigma}}_{eq})}\mathbf{s}
\end{equation}

 In order to calculate a current value of the quantity $S_m$, similarly to the tangent formulation and following \cite{Berbenni21}, we will use the Hill-Mandel's lemmas \eqref{Eq:Hill} and \eqref{Eq:Hill2} with the approximations  applied therein.
 The only difference is in the viscoplastic terms once the secant linearization \eqref{Eq:secant} is applied, namely for the secant model:
 \begin{equation}
 \overline{\boldsymbol{\sigma}\cdot\dot{\boldsymbol{\varepsilon}}^v}_m=\frac{1}{2\mu^{sec}(\bar{\bar{\sigma}}_{eq})}\overline{\mathbf{s}\cdot\mathbf{s}}_m=\frac{1}{2\mu^{sec}(\bar{\bar{\sigma}}_{eq})} S_m
 \end{equation}
and
 \begin{equation}
 \overline{\dot{\boldsymbol{\sigma}}\cdot\dot{\boldsymbol{\varepsilon}}^v}_m=\frac{1}{2\mu^{sec}(\bar{\bar{\sigma}}_{eq})}\overline{\dot{\mathbf{s}}\cdot\mathbf{s}}_m=\frac{1}{4\mu^{sec}(\bar{\bar{\sigma}}_{eq})} \dot{S}_m\,,
 \end{equation}
instead of relations \eqref{Eq:evHillav} and \eqref{Eq:evHill2} valid for the tangent one.
Thus, the final relations for $\dot{S}_m$ and $\ddot{S}_{m}$ are of the same forms: \eqref{Eq:HillResult} and \eqref{Eq:HillResult2} as for the tangent model, however, the approximation \eqref{Eq:ass-cov}$_2$ is here not necessary. 
The solution algorithm for the second-moment (modified) secant formulation is similar as for the modified tangent model, in particular, the quadratic update \eqref{Eq:Sm-update} of $S_m$ is used. 

\begin{figure}
    \centering
        \includegraphics[angle=0,width=0.45\textwidth]{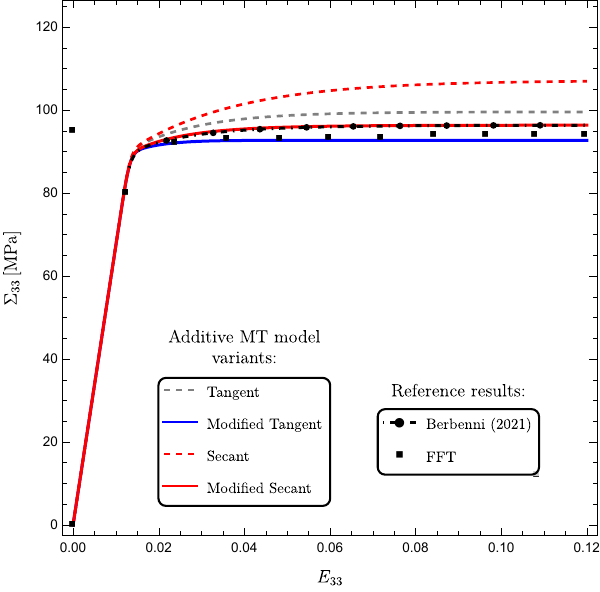}
\caption{Macroscopic axial stress $\Sigma_{33}$ vs. macroscopic axial strain $E_{33}$ during the isochoric tension process \eqref{Eq:isoch} with the strain rate of $0.012$ $[s^{-1}]$. Predictions of the four variants of the additive MT scheme for a two-phase composite with $17\%$ of inclusion are compared with
the FFT results of \cite{Lahellec13} and the secant homogenization model of \cite{Berbenni21}. 
The inclusion has an elastic behavior while the matrix has an elastic-viscoplastic response of the Perzyna-type, see Eq. \eqref{Eq:Perzyna}. Material parameters of both phases are reported in Table \ref{Tab:mater-parameters}.}\label{Fig:SecOverall}
\end{figure}

Figure \ref{Fig:SecOverall} compares predictions of four variants of the additive MT model depending on the applied linearization method. The isochoric tension process with the strain rate of $0.012 [s^{-1}]$ defined in Sec. \ref{Sec:Results} is considered. The evolution of the axial macroscopic stress with the axial strain is shown in Fig. \ref{Fig:SecOverall}. As expected the first-moment secant variant provides the stiffest predictions as compared to other variants. The predictions of the second-moment (modified) secant variant are also stiffer than for the modified tangent model although softer than for the original tangent model. Fig. \ref{Fig:SecOverall} includes also reference results by FFT and predictions of the modified secant mean-field model proposed in \cite{Berbenni21}. It is seen that the stress evolution predicted by this last method is very close to the predictions of the modified secant additive approach.

\begin{figure}
     \begin{tabular}{cc}
     (a)&(b)\\
        \includegraphics[angle=0,width=0.45\textwidth]{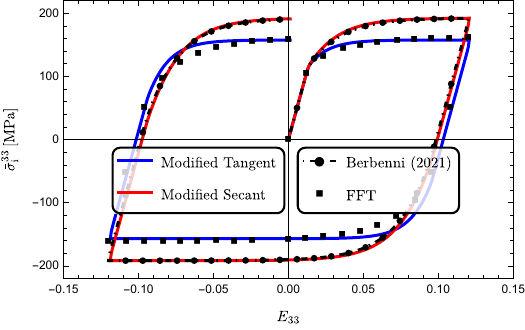}&\includegraphics[angle=0,width=0.45\textwidth]{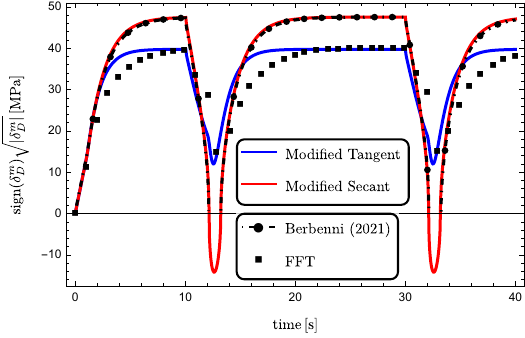}\\
    \end{tabular}
\caption{(a) Mean axial stress $\bar{\sigma}_{33}$ in the inclusion vs. macroscopic axial strain $E_{33}$ and (b) evolution of stress fluctuations in the matrix: $\textrm{sign}(\delta^m_{\rm{D}})\sqrt{|\delta^m_{\rm{D}}|}$ (Eq.~\eqref{Eq:deltaD}) over time. The isochoric tension-compression cycle is prescribed to a two-phase composite with $17\%$ of inclusions. Predictions of the additive modified tangent and modified secant MT schemes are compared with
the FFT results \citep{Lahellec13} and the secant homogenization model of \cite{Berbenni21}. 
The inclusion has an elastic behavior while the matrix has an elastic-viscoplastic response of the Perzyna-type, see Eq. \eqref{Eq:Perzyna}. Material parameters of both phases are reported in Table \ref{Tab:mater-parameters}\label{Fig:InFlu-Sec}}
\end{figure}

Figure \ref{Fig:InFlu-Sec} compares the performance of the variants of the additive scheme based on the second moment of stress in predicting the mean stress evolution in the inclusion phase and the stress fluctuations in the matrix. It is clearly seen that the modified tangent formulation outperforms the modified secant one providing significantly better agreement with the reference FFT results. This is particularly valid for the stress fluctuation. Similarly to the estimates of the overall stress, the predictions of the modified secant additive model are very close  to the results of the model proposed by \cite{Berbenni21}. {It is noted that negative values of $\delta^m_{\rm{D}}$ are observed in the case of modified secant models, which is not observed in FFT calculations.}


\end{document}